\documentclass[useAMS,usenatbib]{mn2e}

\title[Lindblad Zones]
{Lindblad Zones: resonant eccentric orbits to aid bar and spiral formation in galaxy discs}
\author[C. Struck] 
{Curtis Struck \thanks{E-mail: curt@iastate.edu} \\
Department of Physics and Astronomy, Iowa State University, Ames, IA, 50014 USA}
\usepackage{amssymb}
\usepackage{amsmath}
\usepackage{graphicx}
\usepackage{multirow} 
\def\aap{{ A\&A}}

\def\aj{{AJ}}

\def\apj{{ApJ}}

\def\mnras{{MNRAS}}

\begin{document}
\date{\today}

\pagerange{\pageref{firstpage}--\pageref{lastpage}} \pubyear{0000}

\maketitle

\label{firstpage}
\begin{abstract}

The apsidal precession frequency in a fixed gravitational potential increases with the radial range of the orbit (eccentricity). Although the frequency increase is modest it can have important implications for wave dynamics in galaxy discs, which have not been previously explored in detail. One of the most interesting consequences is that for a given pattern frequency, each Lindblad resonance does not exist in isolation, but rather is the parent of a continuous sequence of resonant radii, a Lindblad Zone, with each radius in this zone characterized by a specific eccentricity. In the epicyclic approximation the precession or epicyclic frequency does not depend on epicycle size, and this phenomenon is not captured. A better approximation for eccentric orbits is provided by p-ellipse curves \citep{st06}, which do exhibit this effect. Here the p-ellipse approximation and precession-eccentricity relation are used as tools for finding the resonant radii generated from various Lindblad parent resonances.  Simple, idealized examples, in flat rotation curve and near solid-body discs, are used to show that ensembles of eccentric resonant orbits excited in Lindblad Zones can provide a backbone for generating a variety of (kinematic) bars and spiral waves. In cases balancing radius-dependent circular frequencies and eccentricity-dependent precession, a range of resonant orbits can maintain their form in the pattern frame, and do not wind up. Eccentric resonance orbits require a strong perturbation to excite them, and may be produced mostly in galaxy interactions or by strong internal disturbances. 
 
\end{abstract}


\section{Introduction}

The recognition of bars and spirals as the dominant structure in the `spiral nebulae' came even before these objects were understood as galaxies like the Milky Way (e.g., \citealt{sh43}, \citealt{sh15}). The study of bars and spiral waves in galaxy discs is a significant part of the whole literature of extragalactic astronomy. The majority of this literature is focused on observation analysis and interpretation, or the results of numerical simulations. The simulation literature makes clear that the formation and evolution of these structures is very complex (e.g., recent reviews and texts include: \citealt{se93}, \citealt{bt08}, \citealt{at13}, \citealt{se14}). The simulations have guided us to a fair conceptual understanding of a number of the relevant processes. Examples include the fact that bars can form as a collective response to global, gravitational instability in galaxy discs (e.g., \citealt{mi68}, \citealt{ho69}), and that the self-gravitational effects in `swing amplification' are key to understanding the nonlinear evolution of spirals \citep{to81}. 

The analytic theory of these structures is a smaller part of the literature. There are two core analytic theories: the weak bars theory based on epicyclic orbits (e.g., \citealt{bt08} and \citealt{se93}), and the wave theory of low-amplitude, tightly-wound spirals (\citealt{bt08}). There are strong limitations to the epicyclic approximation to orbits. With a fixed epicyclic frequency the approximation is only accurate for small epicycles, or slightly non-circular orbits. Because of this, the `weak bars' theory might be skeptically viewed as a `broad ovals' theory. However, as \citet{se93} point out, some properties of more eccentric orbits, and thus an extension of the theory, can be obtained from the action-angle formalism (also see \citealt{bt08} and references therein). However, this formalism does not directly yield the shapes of eccentric orbits, which are useful in understanding how they can support various waves. Additionally, there are more complete analytic theories in special cases, such as Freeman's harmonic oscillator potential and the theory of St\"{a}ckel potentials (see \citealt{bt08}). Unfortunately, these potentials are not the most relevant for galaxies. The lack of more comprehensive analytic models for bars has long been noted in the literature. 

\citet{by05} developed an analytic formulation of inner and outer star forming rings in barred disks. The formulation is for a general rotation curve, based on epicyclic perturbations around Lindblad resonances, with terms for weak dissipation.  This formulation and supporting by N-body simulations are compared to Hubble Space Telescope observations of the rings in the galaxy NGC 3081 which has a weak bar and a flat rotation curve.  This galaxy is close enough to see individual young star associations in the rings. The pointed oval shape of the inner (r) ring is produced by the analytic formulation as observed, as are the outer R1 and R2 rings with all turning at the bar pattern speed. The NGC 3081 disk surface density and relative importance of the dark matter halo can be deduced with the formulation.  Some of the eccentric resonant stellar orbits described below resemble the shapes of the star forming rings in the NGC 3081 model. It seems likely that the formalisms are the same in the limits of small eccentricity and weak dissipation.

Recently, in a series of papers, Romero-G\'{o}mez, Athanassoula, and collaborators have presented a theory for rings and spirals associated with bars, which is based on orbits on invariant manifolds in barred potentials. This work, which can be viewed as semi-analytical, is summarized in the review \citet{ro12}. It is supported by N-body results  \citep{at12}. Some of the eccentric, resonant orbits described below have a similar appearance to the invariant manifold orbits, and can account for similar observed morphologies in galaxies. This is remarkable since only axisymmetric potentials are used in the present work. It may suggest that the invariant manifolds, as well as the bars themselves, are seeded by eccentric resonant orbits. 

The literature on tidally induced bars and spirals in tidal interactions is also very modest, and it is dominantly focused on the results of numerical simulation (e.g., \citealt{no87}, \citealt{no88}, \citealt{ge90}, \citealt{mi98}, \citealt{be04}, \citealt{yo14}, \citealt{lo14}). In many ways the analytic theory of the general case can be carried over to the tidal case, because the general theory is modal, and the lowest order even mode (`m = 2') is the most relevant in both cases. However, tidal waves produced in galaxy interactions are not generally weak, nor of low amplitude. It seems fair to say that at present there is little or no specific analytic theory for them (but see a special case in \citealt{st11}, \citealt{do14}). The formalism presented below provides some insight on what perturbation amplitudes are required to excite orbits that might support the tidal waves.

In the following sections resonant eccentric orbits which may help fill several gaps in the basic theory of bars and spiral waves are described and categorized. Firstly, they offer the chance to extend models based on simple epicyclic orbits, helping to bridge the gap noted above between such models and numerical simulations. They also can provide considerable guidance to the study of orbits in developing waves in simulations. Beyond this their existence, shapes, and systematics can provide a broader conceptual framework for understanding non-axisymmetric waves than the near circular epicyclic formalism. One of the keys to achieving these goals is a better modeling of the precession of eccentric orbits, how it relates to orbital resonance. We conclude this section with a brief introduction to that subject. 

In a galaxy disc a wave with fixed pattern frequency, $\Omega_p$, is in resonance with near circular orbits where the Lindblad condition is satisfied,

\begin{equation}
\label{eq1}
{\Omega}(r_o) \pm \frac{{\kappa}(r_o)}{n_r} ={\Omega}_p,
\end{equation}

\noindent
where $\Omega$ is the circular frequency at radius $r_o$ in the galactic gravitational potential, and $\kappa$ is the epicyclic frequency. It is assumed that the near circular orbit is well approximated by a circular epicycle rotating around a circular guiding centre. When equation \eqref{eq1} is satisfied the orbit is closed, and if $n_r = 2$ it is a symmetric ellipse (an inner or outer Lindblad resonance, ILR or OLR, see \citealt{bt08}).  

The epicyclic frequency gives a simple approximation to the orbital precession frequency, which is independent of the size of the epicycle. In fact, the precession frequency of an orbit in a general potential depends on the radial excursion of the orbit. The precession frequency increases with the magnitude of the radial excursion, or the appropriately defined eccentricity. For general orbits, equation \eqref{eq1} can be replaced with,

\begin{equation}
\label{eq2}
\left[ 1 \pm \frac{m(e)}{n_r} \right] \Omega
= {\Omega}_p,
\end{equation}

\noindent
where $m$ is the ratio of the precession to the orbital frequency. If the pattern speed is held constant, as $m$ exceeds $m(e=0) = {\kappa}/{\Omega}$, then equation \eqref{eq2} will not continue to hold unless ${\Omega}(r)$ is adjusted. As $\Omega$ generally varies with radius, this implies that the resonant radius corresponding to an orbit with a given eccentricity, $e$, is different than the radius of the near circular Lindblad resonance ($r_o$). Exceptions include the solid body potential where $\Omega$ is constant, and cannot be adjusted, and the point-mass potential where there is no precession. 

Aside from the exceptional cases, these considerations imply the existence of different resonant radii for every value of the eccentricity, when the latter is not negligible. As will be shown below, the differences between the resonant initial radii of eccentric orbits and the parent Lindblad resonant radius can be significant, though not generally large. Thus, for a given pattern speed the discrete (near circular) Lindblad resonance radii spawn a continuum of eccentric resonant radii, which define the `Lindblad Zones' of the title. An example of the usual diagram of  frequency versus radius in a flat rotation curve case, with the addition of sample eccentric resonant frequencies, is shown in Fig. 1 below. We consider some specific examples of such orbits and these frequencies in the following.

\section{Applying p-ellipse approximations to eccentric resonances}
\subsection{Review of p-ellipse orbit approximations}

For orbits of moderate radial ranges in single power-law potentials, approximations of the form,

\begin{equation}
\label{eq3}
\frac{1}{r} = \frac{1}{p} \left[ 1 +
e \cos \left( m{\phi} \right) 
\right]^{\frac{1}{2} + \delta},
\end{equation}

\noindent were described in \citet[Paper 1]{st06}, named `p-ellipses', and found to be quite accurate (for other approximations see \citealt{va12}). Here the orbital scale is given by the semi-latus rectum $p$, the azimuthal angle is $\phi$, and $e$ is the eccentricity parameter. The gravitational acceleration is taken to be of the form,

\begin{equation}
\label{eq4}
g_r = -\frac{GM}{r^{2\delta + 1}},
\end{equation}

\noindent
with the radial dependence specified by the exponent $\delta$. The mass $M_{\epsilon}$ is the mass contained within a scale radius $\epsilon$. Rather than use the equation of motion for $r$ in terms of time $t$, it is more convenient to use the dimensionless equation for $u = {\epsilon}/r$, as a function of the azimuth $\phi$. This transformation, described in Paper 1, yields the equivalent equation of motion,  

\begin{equation}
\label{eq5}
u'' =  {c_\delta}u^{{2\delta}-1} - u,
\end{equation}

\noindent where the double prime notation indicates the second derivative with respect to $\phi$.

In dimensional units,

\begin{equation}
\label{eq6}
c_\delta = \frac{\epsilon}{p^{2\left(1-\delta\right)}},
\end{equation}

\noindent where this equality is derived using equation (B1) of Paper 1, relating the specific angular momentum $h$ to scale size $p$ of a p-ellipse orbit. Dimensionless units are defined by setting $\epsilon = 1$, $GM{\epsilon} = 1$, and a time unit of,

\begin{equation}
\label{eq7}
\tau = \left( \frac{GM_{\epsilon}}{{\epsilon}^3} \right) ^ {-1/2}.
\end{equation}

\noindent An equation for the azimuthal advance derives from specific angular momentum conservation,

\begin{equation}
\label{eq8}
\frac{d{\phi}}{dt} = \frac{h}{r^2} =
\frac{h}{p^2} \left[ 1 + e\ cos(m{\phi}) \right] ^{1+2{\delta}}
\end{equation}

\noindent where the second equality is the p-ellipse form obtained by substituting equation \eqref{eq1}. 

In the present work we will focus on two particular potentials, a perfectly flat rotation curve (FRC) potential ($\delta = 0$) and a near solid-body potential ($\delta \simeq -1$). In the former case equation \eqref{eq8} is integrable, and the orbital period $T$ can be derived as a function of the orbit parameters, see Appendix C of Paper 1. Specifically,  

\begin{equation}
\label{eq9}
T = \frac{p^2}{mh} \frac{2 \pi}{\sqrt{1-e^2}}.
\end{equation}

\noindent In \citet[Paper 2]{st15} the p-ellipse approximation was improved and extended.  A 3-point or three parameter elaboration of equation \eqref{eq3}, which included a single additional harmonic term was found to provide very accurate approximations to orbits with very large radial ranges.  However, the 3-point equations are more complex, and the extra dimension makes the parameter space much larger. The essential points of the present work can be illustrated with the simpler approximation of equation \eqref{eq3}, so that approximation, with some improvements, will be used here.

The more accurate approximations of Paper 2 were based on three developments. The first was to determine the p-ellipse parameters in orbit fitting using the minimum and maximum radii, rather than for example the radius and velocity at a single position. This guarantees that approximate and true orbits traverse the same radial range. This improvement was discovered in a similar context by \citet{ly10}. The second was to determine the dependence of the parameter $m$ on the p-ellipse eccentricity, $e$ (also see \citealt{va05}, \citealt{ly15}). This keeps the approximate and true orbits of high eccentricity in phase. The third improvement, adding the harmonic term, allows the approximate orbit to better fit the shape of the true orbit. Since fitting true orbital shapes and ranges is not a major concern of this work, we forego this improvement, and the complexity introduced by the first. Accurate precession frequencies are crucial however, so the $m(e)$ correction must be included. 

\subsection{Exciting resonances in limited disc regions}

In order to illustrate the possible roles of eccentric resonant orbits in disc waves, some relatively simple examples will be presented in the following sections. These are usually based on the premise of exciting a small region of the disc, with little or no excitation in surrounding regions. This is certainly plausible if the excitation source is something like a large clump in a young disc, or perhaps, a collision with a dwarf companion galaxy. Plausible, at least, for such sources acting on a relatively short timescale; clumps may have global effects on longer timescales, e.g.,  \citep{bo07}.

On the other hand, the most important disturbances of galaxy discs, such as massive interactions or accretion events, have global effects. However, even these events will not generally effect all parts of the disc equally, and if impulsive, may be focused on one region. A prototypical case is an impulsive tidal interaction, where the strongest torques are exerted along disc longitude lines offset by about $45^{\circ}$ from the line connecting the disc centre to the companion galaxy (see e.g., \citealt{st99}). disc regions nearer to the companion will receive a primarily radial perturbation.

In the Impulse Approximation, radial velocity perturbations generally have smaller effects than azimuthal perturbations from a circular orbit. In the case of a purely radial disturbance, comparing the energy equation both of an initially circular orbit (e.g., in a FRC potential) immediately after the impulse and at greatest radial excursion, one obtains, ${\Delta}r/r \simeq ({v_r}/{v_\phi})^2$, where $v_\phi$ is the original circular velocity. In the case of a purely azimuthal disturbance, a similar comparison of angular momenta yields, ${\Delta}r/r \simeq {{\Delta}v_{\phi}}/{v_\phi}$. If the two velocity impulses ($v_r$ and $\Delta v_{\phi}$) are roughly comparable, and both are less than the circular velocity, $v_{\phi}$, then the radial perturbation is quadratic in a small velocity ratio, while the azimuthal is linear. Thus, the highest magnitude disturbance is along a limited region around the $45^{\circ}$ longitude lines. 

A specific example will further emphasize this point. In some impulsive FRC examples below, a moderate pattern frequency was assumed, about $1/3$ of the circular frequency, or of the circular velocity at $r = 1$ in dimensionless units. A frequency value of, e.g., a factor of 2 higher or lower could have been used, but in the FRC case the ratio of pattern to circular velocity at resonances would not change much. The chosen value requires a substantial disturbance to impulsively excite orbits in a patch in the inner Lindblad Zone. The velocity impulse is roughly $\Delta v = G{M_p}{\Delta}t/{r_p}^2$ for a perturbing mass $M_p$ at a distance of $r_p$ acting for a time ${\Delta}t$. If, for example, we require a velocity reduction of $100\  km\ s^{-1}$, and assume the impulse acts entirely in the azimuthal direction, we can write,

\begin{equation}
\delta{t} =2.3 \times 10^8\ \left( \frac{10^{10} M_\odot}{M_p} \right)
\left( \frac{r_p}{10\ kpc} \right) ^2
\left( \frac{{\Delta}v}{100\ km\ s^{-1}} \right) \ yr.
\label{eq15}
\end{equation}

\noindent A large molecular cloud or globular cluster would not provide a sufficient force for an impulse of this magnitude. It would take an intermediate mass companion galaxy or a substantial non-axisymmetric wave or clump within the disc to provide a sufficient pull within a time less than a typical orbital time. Thus, exciting very eccentric resonant orbits in the Lindblad Zone is relatively difficult. However, given such as disturbance, the direct formation of a wave backbone consisting of a set of eccentric orbits with correlated phases is possible. Since these orbits will generally derive from a small fraction of the disc, only a small fraction will be very eccentric. 
 
\section{Exciting a Lindblad Zone to aid bar formation}
\subsection{Sample ILR orbits in the flat rotation curve case}

In this section we use the formalism of the previous section to derive specific examples the eccentric resonant orbits mentioned in the Introduction. Both p-ellipse approximations and numerical integrations will be used, the latter to confirm the validity of the approximate results. In the FRC case, the precession frequency parameter is well approximated as (see Paper 2),

\begin{multline*}
\frac{m(e)}{m_o} = 1.0013 - 0.00439x + 0.0520x^2
+ 0.0169x^3\\ + 0.00180x^4,
\end{multline*}
\begin{equation}
\label{eq10}
m_o = \sqrt{2(1-{\delta}}, \ \ x = log_{10}\left( 1-e \right).
\end{equation}

\noindent Clearly, for small values of $e$, $x$ is also very small, and the right hand side of the first of these equations is nearly $1.0$. Even at large eccentricities the ratio $m(e)/m_o$ does not differ greatly from unity. For example, when $e = 0.98$, the value of this ratio is $1.0910$. (The $m(e)$ relation was also described, though not with the $ln(1-e)$ scaling in \citet{va97}, \citet{va99} and \citet{va05}. The dependence on different potentials and specific angular momentum were also described in these papers.) 

A side note, because of the nonlinear exponent in equation \eqref{eq3}, the p-ellipse equation does not give the same value of the maximum to minimum radius ratio, $(1+e)/(1-e)$, as for simple ellipses. The corresponding ratio for p-ellipses is $[(1+e)/(1-e)]^{1/2+\delta}$.  For $e = 0.9$, this yields a factor of $4.36$, not $19$ as in the case of simple ellipses. Thus, fairly large values of the p-ellipse $e$ do not necessarily imply equally extreme radial orbits in the FRC case. 

For each value of $m(e)$, there is an $\Omega_1$, given by equation \eqref{eq2} for the resonant orbit of that eccentricity. (The subscript `$1$' is henceforth adopted for eccentric orbits and the subscript `o' for circular orbits, especially that of a parent Lindblad resonance.) We use the definition of $T$, and equation \eqref{eq9} to obtain another expression for $\Omega_1$, 

\begin{equation}
\label{eq11}
\Omega_1 = \frac{2{\pi}{\tau}}{m_1(e)T}
= \frac{{\epsilon} \sqrt{1-e^2}}{p} ,
\end{equation}

\noindent where we include the units $\epsilon, \tau$, which equal $1.0$ in the dimensionless system. By combining equations \eqref{eq2} and \eqref{eq11} we can solve for $\Omega_1$ and $p$ in terms of $e$, thus, completely specifying the approximate resonant orbit for a given $e$ and pattern speed. Specifically, for $p$ we have, 

\begin{equation}
\label{eq12}
\frac{p}{\epsilon} = \left( \frac{1 - \frac{m_1}{2}}{\Omega_p} \right)
\sqrt{1-e^2} = \left( \frac{1 - \frac{m_1}{2}}{1 - \frac{m_o}{2}} \right)
\sqrt{1-e^2},
\end{equation}

\noindent where with the choice of the minus sign in the brackets and $n_r = 2$, we have also specialized to the case of the inner Lindblad resonance (ILR). In the last equality of equation \eqref{eq12} we have chosen to set the pattern frequency equal to $1 - {m_o}/2$, which has the advantage of putting the (circular) ILR at $r_o = 1$. This choice will be used throughout this work.

These formulae for $p$ and $m_1$ can then be used in equation \eqref{eq3} to determine the minimum and maximum radii of the orbit. Finally, by comparing to the ILR orbit, we obtain an expression for the radius of origin, $r_1(e)$, of the resonant eccentric orbit. I.e., the radius of the circular parent orbit of the eccentric orbit, analogous to the guiding centre of an epicyclic orbit,
\begin{equation}
\label{eq13}
\frac{r_1(e)}{\epsilon} = \frac{\Omega_\circ}{\Omega_1}.
\end{equation}

\noindent The value of this radius of origin parameter was used to produce the dashed curves in Fig. 1 for the ILR, and in the same manner for the outer Lindblad resonance (OLR), and the $n_r = 1$ case shown. 

\begin{figure}
\centerline{
\includegraphics[scale=0.35]{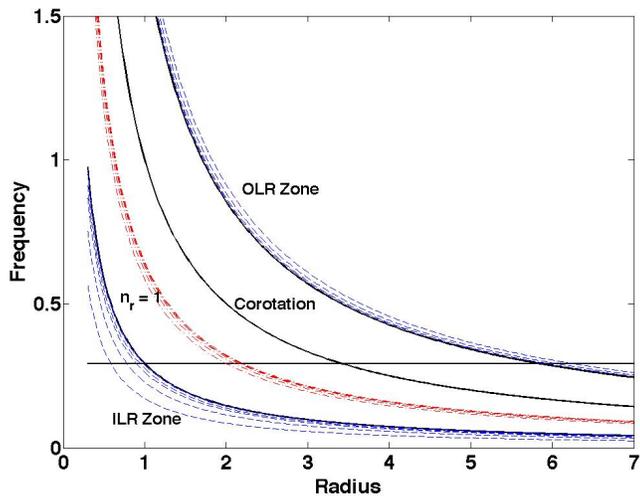}}
\caption{Resonant frequencies versus radius in a flat rotation curve potential. The flat horizontal line gives the pattern frequency. The black solid lines are, from the lowest frequency to the highest at a given radius: the Inner Lindblad resonance, corotation resonance, and the outer Lindblad resonance for the given pattern frequency. The dashed blue curves show the eccentric resonances, and highlight the Lindblad Zones containing them, for the ILR and OLR the eccentricity $e$ values are $(0.3, 0.5, 0.8, 0.9, 0.98, 0.999)$. The red, dashed curves are the eccentric resonances for the $n_r = 1$ case, with the same eccentricities. Radii in this and all the following figures are in units of a scale length $\epsilon$, and times or inverse frequencies in units of  $(GM_{\epsilon}/\epsilon^3)^{1/2}$, where $M_{\epsilon}$ is the mass contained within the scale length. }
\end{figure}

Figure 2 shows 5 examples of approximate resonant eccentric orbits over a range of eccentricity, in a reference frame rotating at the pattern frequency, computed with equations \eqref{eq3} and \eqref{eq10}-\eqref{eq12}. The initial condition in each case was to start on the x-axis with the minimum radius. A final panel of the figure shows the fifth orbit again, but in the rest frame, rather than in the rotating frame. As in that sixth panel, these orbits do not generally close in the non-rotating reference frame. They look like the p-ellipse examples of Papers 1 and 2. One feature is immediately clear from this figure, except for examples with the lowest eccentricities, the orbital shapes differ noticeably from the simple ellipses. (They are, however, like the periodic orbits found in numerical simulations of bars.) At moderate eccentricities they have cusp-like ends, and extra loops at high eccentricities. The minor axis size of these orbits decreases with increasing eccentricity making thinner forms. The major axis changes little with eccentricity, making the major axis length similar in all cases. This is because the factor $p$ in equation  \eqref{eq3} decreases almost inversely with the $e$ term. Thus, at maximum radii $r = p/(1-e)^{1/2}$ these two dependences roughly cancel. At minimum radii they reinforce each other. These trends are confirmed in numerically integrated orbits described below; the analytic approximation is helpful for understanding why they occur.

\begin{figure}
\centerline{
\includegraphics[scale=0.40]{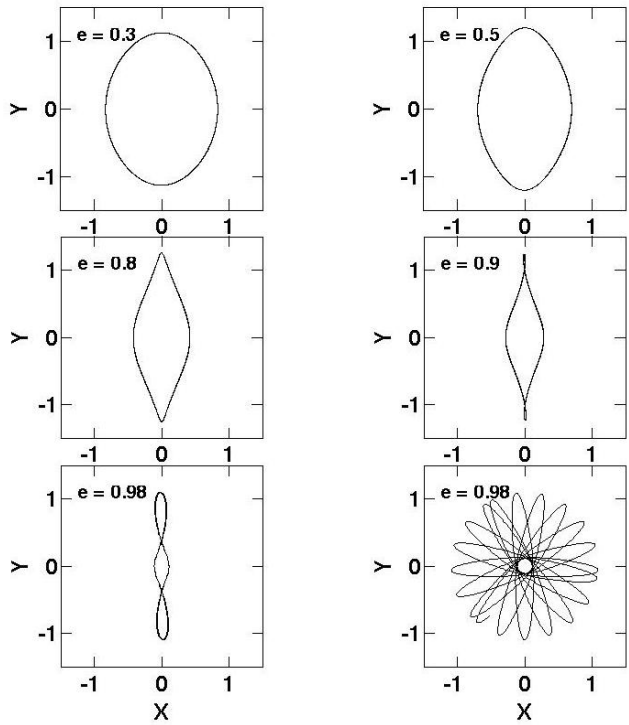}}
\caption{Sample, analytic resonant orbits in a flat rotation curve potential derived from equations \eqref{eq3} and \eqref{eq10}-\eqref{eq12} in the inner, $n_r =2$, Lindblad Zone. These are shown in the reference frame rotating with the pattern frequency, except for the lower right panel. That orbit is the same as the one in the lower left panel, except shown in the non-rotating frame. The eccentricities are labeled in each case. }
\end{figure}

A stellar bar could certainly be made from the in-phase superposition of the more eccentric of these orbits. Specifically, via the mechanism of \citet{ly79}, or of \citet{po89} if enough eccentric orbits were excited. In all cases the outer radii of these orbits are less than about a factor of $1.25$ times the radius of the (circular) ILR. This is partially a function of the initial conditions. Their inner radii can be arbitrarily small, but the smallest occur only with eccentricities very close to $1.0$. Since all of these orbits are closed in the pattern frame, a bar made of them is permanent in the limit of no disturbances. 

The possibility of stable, non-self-gravitating bars in a symmetric potential is quite interesting. In principle, these orbits differ from closed orbits in imposed bar-like potentials, which are often proposed as the parents of classes of orbits sustaining the bars (\citealt{bt08}, \citealt{se93}). In fact the two classes may be closely related. \citet{ly96}, in a general discussion of bar making, describes a problem with his \citep{ly79} model, ``strong bars often rotate so fast that no inner Lindblad resonances exist within the bars themselves.'' Highly eccentric ILRs, may do so, and persist in self-gravitating bars (see Fig. 1). 

The shape of a bar formed from these orbits would clearly depend on the range of eccentricities included. Wide bars or oval distortions could be formed from low-$e$ orbits, narrow bars from high-$e$ orbits. Thus, depending on what range of eccentric resonances are excited by a disturbance, a range of bar shapes could be produced with these orbits. 

The fact that these results all derive from an approximate analytic solution may raise the concern that perhaps they are artifacts of the approximation. Figure 3 shows four orbits obtained by numerically integrating equation \eqref{eq5} with initial conditions corresponding to those used to produce the analytic orbits of Fig. 2. Specifically, the initial position was a minimum radius like the corresponding analytic model, where the radial velocity is zero. Additionally, the force parameter $c_\delta$ is determined by the corresponding value of $p$. Using the exact same minimum radius values as in the corresponding analytic case does not generally yield closed or nearly closed orbits like those in Fig. 3. Given the inaccuracy of the approximation at the level used above, as described in Papers 1 and 2, this is not surprising. The curves of Fig. 3 were obtained by experimentally adjusting the value of the initial radius until near-closure was obtained. 

\begin{figure}
\centerline{
\includegraphics[scale=0.35]{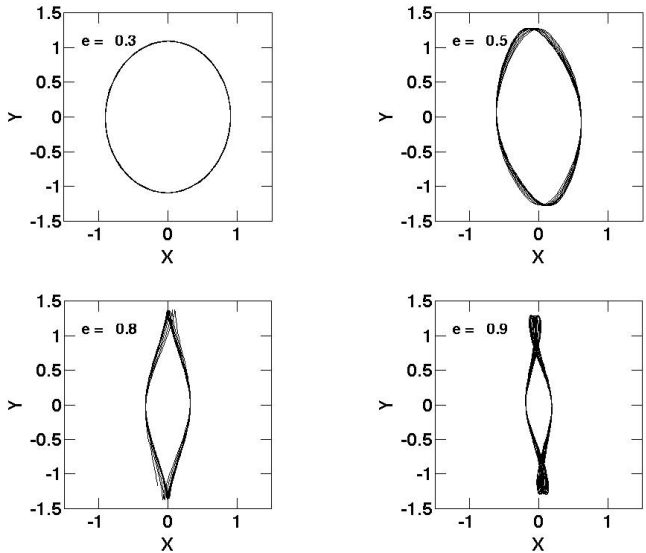}}
\caption{Numerically computed near resonant orbits, like the analytic orbits of Fig. 2. Eccentricities of the corresponding analytic orbits are labeled on each panel, the initial radii are $(0.905, 0.6127, 0.3242689, 0.18914232)$.}
\end{figure}

This procedure also yielded valuable insights about nearby orbits. Firstly, except at small eccentricities it is difficult to determine an exact value of $p$ that yields a perfectly closed orbit. (One minor aspect of this is that the integrator used, the ODE45 code of the Matlab software package, is Runge-Kutta based, not sympletic, so its errors are expected to grow, and convergence cannot be checked over a large number of orbital periods.) Secondly, in a modest range of initial radius values there are many sub-ranges containing nearly closed orbits, interspersed between sub-ranges occupied by very open orbits in pattern frame. While this makes it difficult to find the closed orbit, it has the positive benefit, from the point-of-view of bar making, that each closed orbit has many nearly closed sister orbits. Finally, none of these caveats changes the conclusion that each approximate orbit of Fig. 2 has a corresponding `true' orbit that can be found (or at least narrowly bracketed) numerically. This justifies the further use of the p-ellipse toolkit. 

\subsection{Orbits near other resonances}

The next resonance to consider is the OLR. As can be seen in Fig. 7 below, the shape of the eccentric OLR orbits is like that of a double Lima\c{c}on, and they are generally large orbits. Actually, it is a pseudo-Lima\c{c}on, since the equation has an exponent not present in that of the Lima\c{c}on. The outer part of the curve is close to circular, but the opposing inner `pedals' could contribute to a bar. However, despite having the same initial conditions as Fig. 2, these pedals line up along the x-axis, rather than the y-axis. Moreover, almost all of these OLR orbits lie outside the ILR, so a bar made of orbits from Fig. 2 would lie completely interior to one consisting of OLR orbits. As discussed below, these OLR orbits seem better suited to making rings or spiral waves. 

Next we consider co-rotation resonance. The orbits associated with co-rotation are ovals around the point at the co-rotation radius where they are excited. Because, $n_r = 0$, in this case, the eccentricity dependence of the precession does not matter, and the Lindblad zone collapses to the co-rotation radius itself. All elliptical orbits originate from that radius. If a range of eccentricities are excited, then orbits form a set of concentric (but not similarly shaped) ovals around the excitation point. If a range of azimuths are excited, we will have a range of offset ovals of different sizes. It is hard to imagine making bars or spirals from an ensemble of these orbits, but it is easy to make arcs (or the 'ansae', see \citealt{bu13}), or potentially rings, like those often visible outside of (ILR) bars. 

Then come the eccentric $n_r = \pm1$ cases. The shapes of the resonant orbits in these cases are similar to those in the ILR and OLR cases. With an orbital frequency obtained by solving $(1 - m_1)\Omega_1 = \Omega_p$, this case has nearly elliptical orbits at all eccentricities. However, each of these ellipses is offset from the disc centre. At high eccentricity it can develop a cusp at the small radius point on the major axis, see the lower panel of Fig. 7 below. $m_1(e)$ is generally greater than unity, so in this case we choose the negative value of the square root term in equation \eqref{eq12} to get a positive value of $p$. This yields a negative value of $\Omega_1$, and these stars orbit in the opposite sense of previous ones. An ensemble of such orbits could make up an asymmetric bar, or if excited symmetrically on both sides of the disc, a symmetric bar, along the x-axis. 

In the case of an orbital frequency determined by $(1 + m_1)\Omega_1 = \Omega_p$, we get pseudo-lima\c{c}on orbit shapes as in the case of the OLR eccentric family. Inner pedals of these orbital curves could contribute to bars in the same way as the previous case.

Fig. 4 shows examples of four $n_r = 4$ orbits, which have four maxima or minima in each orbital circuit in the pattern frame. These orbits are also members of a Lindblad Zone. The symmetries of these orbits make it clear that they would not support a bar, but their shapes make them interesting curiosities. This is especially true in light of the existence of apparently square or rectangular galaxies in a couple of interacting systems, e.g., Arp 25 \citep{ar66} and LEDA 074886 \citep{gr12}. Fig. 5 shows a couple more harmonics, with $n_r = 3$ and $5/2$. These orbits are still quite symmetric, and unlikely to contribute much to a bar potential. A similar result is found for subharmonic orbits with $n_r = 1/2$. Moreover, these orbits tend to have large values of $p$, which translates into large orbital radii.

\begin{figure}
\centerline{
\includegraphics[scale=0.35]{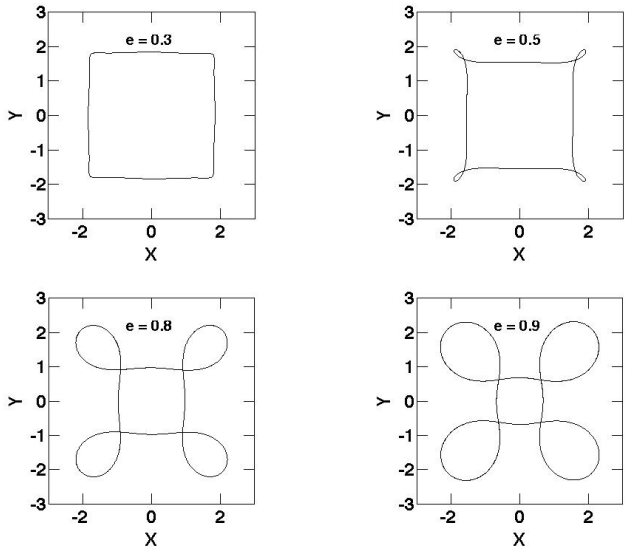}}
\caption{Examples of analytic, harmonic resonant orbits with $n_r = 4$, and different eccentricities, as labeled. See text for further details. }
\end{figure}

\begin{figure}
\centerline{
\includegraphics[scale=0.3]{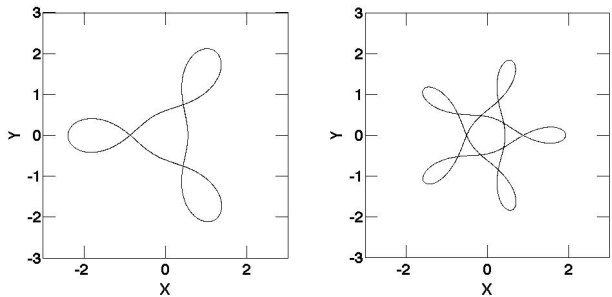}}
\caption{More examples of analytic, harmonic resonant orbits, but with uneven harmonics. The curve in the left panel has $n_r = 3, e = 0.9$, while that in the right panel has $n_r = 5/2, e = 0.9$. }
\end{figure} 

\subsection{Binding resonant orbits for weak bar formation}

Verifying that approximate analytic resonant orbits also appear as numerical solutions to the nonlinear equation of motion, as in Sec. 3.1, is one step toward affirming that such orbits might play a role in galaxy discs. A second issue is how robust these orbits are, for example, do they continue to exist in the presence of a small perturbing force. This force might be due to either the local self-gravity of an ensemble of resonant waves, or an external source. 

A hint of the answer to this second issue is found in numerically derived orbits in weakly barred potentials that look very much like those of Fig. 3. See for example, Sec. 3.3 in \citet{bt08}, especially Figs. 3.15 - 3.19. 

As a further check, a number of calculations were carried out in which a small perturbing potential was added to the code used to produce the orbits of Fig. 3. The perturbing potential was of harmonic oscillator form, with the force aligned perpendicular to the (rotating) bar. For very small magnitudes of the perturbing potential, up to about 1\% of the gravitational potential at the initial radius, there was no great change in the orbits of Fig. 3. The initial radius of the orbit only had to be adjusted slightly, e.g., up to a few percent for a 1\% perturbation, to get the same closed or nearly closed orbit.

Such a simple adjustment is not sufficient to maintain similar closed orbits for perturbations at the level of 1-10\%. An additional, moderate (up to 30\%) change of the $m_1(e)$ is also required. That is, at this level the perturbing force is strong enough to have an effect on the precession rate, and that factor must be accounted for. Once this is done, closed orbits very similar to those with no perturbing force are obtained, and the issue is resolved. 

With these results we can describe a qualitative scenario for weak bar formation with resonant eccentric orbits, which parallels the usual discussion in terms of epicycles. The process begins with the excitation of a group of resonant orbits. As a concrete example consider an impulsive disturbance that excites orbits of the Inner Lindblad Resonance zone. Specifically, suppose that the disturbance changes the angular momentum of orbits in some part of the ILR zone such that they are now closed orbits at a particular pattern frequency. Note that this pattern frequency is set by the precession frequency, or $m_1(e)$, with $e$ corresponding to the new angular momentum of stars near the ILR. It is not determined, for example, by the orbit frequency of the perturbing source. 

For a moderate external perturbation, the disturbance will be small or primarily radial around most of the circle at the ILR radius. At such azimuths the perturbation will result epicyclic orbits. Only at the azimuths where the torque and consequent angular momentum change are relatively large will significantly eccentric orbits be generated. Thus, we expect that these orbits will be derived from a relatively small patch on the disc, and consist of a small fraction of the disc stars (see Sec. 2.2). 

However, whatever slice of the Lindblad zone provides these resonant orbits, we expect that adjacent slices, with similar precession frequencies and angular momentum changes, will put stars on similar, nearly closed orbits. These stars will be part of the bar initially, and take some time to drift away from it, unless bonded by self-gravity. In very fortuitous cases, the angular momentum changes will vary like $m_1$ throughout the Lindblad Zone, and a large part of the zone will excite eccentric orbits. If the disturbance has tidal symmetry resonant orbits will be excited on opposite sides of the discs, and there will be many more stars involved.  It seems plausible that the incipient bar will possess some self-gravity.

Gas clouds may help it acquire more self-gravity. Clouds in the excitation patch will initially follow the stellar orbits. As they move inward, they will be compressed, and collide with clouds on circular orbits. This may add considerably more gas mass to the bar potential, at least temporarily. If these compressed clouds form stars, these stars may also find themselves on bar-like orbits and remain in the bar potential longer than their parent gas clouds. 

If the bar acquires some self-gravity it may excite other resonances like the $n_r = 1$, OLR, or corotation resonances. However, these orbits are elongated along a line perpendicular to the ILR associated orbits, and so, could oppose the buildup of the incipient bar, though they are generally found at larger radii. 

At this point, the process may have produced either a transient tidal bar, or perhaps, a more long-lived weak bar. We will  consider a simpler case in Sec. 5. 

It is worth recalling that in the limit of small eccentricities the p-ellipse approximation is equivalent to the epicyclic approximation, and is a significant extension of it at larger eccentricities.  This extension may be very important for the theory of weak bars developed in the epicyclic approximation, which is described in Sec. 3.3 of \citet{bt08}. p-ellipses better approximate eccentric, looping orbits, and could extend the theory beyond near circular, epicyclic orbits to ones like those in Fig. 2.  This would certainly help bridge the gap between nearly circular orbits, and those supposed to support a bar. 

There is an apparent contradiction between the results above and the epicyclic theory, however. In Sec. 6.5.2 of \citet{bt08}, it is shown that in the epicyclic theory the bar orbits must lie between the ILR and corotation radii. Orbits beyond the OLR are also allowed in the model, but seem unlikely physically. On the other hand, it was shown above that the orbits of Fig. 2, originate in the Lindblad Zone, which lies interior to the classical ILR radius. The resolution to this paradox is that eccentric orbits originate at their individual resonances in the Lindblad Zone. 

Orbital rearrangement is roughly a zero sum game, stars are taken out of the symmetric part of the potential and, in the case of very eccentric ones, added to the bar part. Stars orbiting in the bar will see a weaker symmetric potential, and if they are orbiting within the depths of the bar potential they will not feel most of its gravity either. Thus, their orbital and precession frequencies will be reduced, as will the pattern speed. This latter change will move the ILR outward, and in consequence the bar may growth (in size as well as strength). Until it is supported by careful N-body simulations, this growth scenario is speculative, but suggests interesting directions for future work. 

\section{Single Lindblad Zone spirals}
\subsection{Inner spirals generated from the ILR}

The bar-like orbits of Fig. 2 are all initialized at the same phase value (and position), and in discussing bar creation it was assumed that they were excited from a small range of radii. It is interesting to consider a different limiting case. That is, when resonant orbits are excited at a range of eccentricities across the Lindblad Zone, and each at a slightly different azimuth. This is admittedly a fine-tuned initial condition, i.e., a perturbation that changes the stellar angular momentum by a rapidly increasing amount across a narrow Lindblad Zone. It is an idealized example designed to explore extreme effects within the zone. 

Two examples are shown in the pattern frame in Fig. 6. In both cases orbits are shown with $e$ values ranging from $0.40$ in steps of $0.01$ up to a maximum of $0.96$. In the top panel each successive orbit is offset in azimuth from its predecessor by an arbitrary angle of $\pi/24$, while in the bottom panel the offset is $\pi/40$. This is the only difference between the two examples. 

\begin{figure}
\centerline{
\includegraphics[scale=0.5]{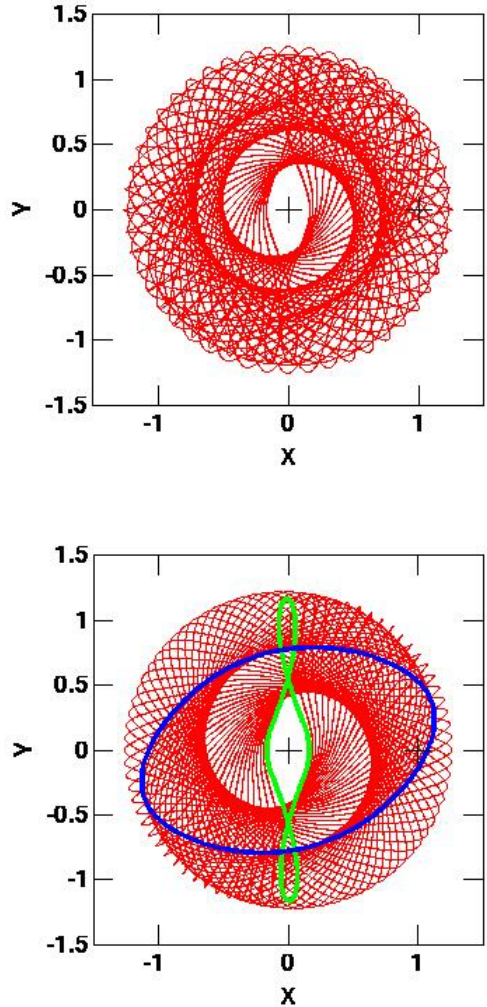}}
\caption{Two examples of spiral waves derived from sequentially offset resonant orbits in the ILR zone. In both cases the eccentricities of the orbits range from $0.4$ to $0.96$ with increments of $0.01$. The successive azimuthal offsets are $\pi/24$ and $\pi/40$ for the top and bottom panels, respectively. In the bottom panel, the bold blue curve highlights the low eccentricity ($e = 0.4$), orbit, while the bold, green curve shows the high eccentricity ($e = 0.96$) orbit.}
\end{figure}

The result is much like the classical twisted ellipses diagram used in textbooks to illustrate how (epicyclic) elliptical orbits may produce spiral waves (see \citealt{bt08}, Fig. 6.12, and \citealt{ka73}). Indeed, very nice spiral patterns are evident in Fig. 6. The bottom panel highlights the orbits with the smallest and largest values of $e$. It can be seen that the former fits onto the spiral between phases of minimum and maximum radius exactly like the twisted ellipse diagram. On the contrary, the most eccentric orbit only does so on its inner loop. Its outer loop cuts the spirals. In the top panel this spiral cutting is even more prominent with the greater offsets between orbits. 

This behavior is novel compared to the usual pictures of how stellar orbits interact with spiral waves. The twisted ellipse picture is the norm for weak spirals. A picture of orbits that intersect a wave, are gravitationally bent along the wave, is often given in the case of stronger waves. The very eccentric orbits of Fig. 6 both follow and cut the kinematic wave at different phases. A self-gravitating wave would have to be quite strong to bend them at either phase. This is because when they align with the wave their orbital momentum is large, but when they cross, they do so at a large angle. 

The classical twisted ellipses are used to illustrate another facet of spiral waves - wind-up. The ellipses continue to twist due to differential precession, or equivalently the radial dependence of the epicyclic frequency, and this drives wind-up. This is not the case for the examples in Fig. 6, like the resonant bar orbits above, they and the patterns they produce are permanent in the pattern frame, subject to self-gravitational or external disturbances. Moreover, unlike the eccentric bar orbits, which were suggested to derive initially from a patch in the disc, those of Fig. 6 originate in a curved region across the Lindblad Zone. Thus, they may involve a greater fraction of orbits than in a single patch, at least initially. 

For any gas clouds excited onto eccentric orbits, the combination of parallel and crossing orbits should have various interesting consequences for the buildup versus disruption of cloud complexes in spiral arms, and perhaps, the development of feathers or spurs from spiral arms. However, the gas cloud orbits could probably not survive multiple arm crossings at high relative velocities, and could not persist on very eccentric orbits. 

The nature of the spirals in Fig. 6 is very interesting. These arms are partially material features. That is, they are produced by a distinct fraction of disc stars, that spend a part of their orbital cycle within the arm, and could do so indefinitely. However, unlike the simplest case of a material arm, these stars are not entirely confined to the arms. All the other stars at these radii would see the spiral as a wave, which they pass through periodically. If that wave develops strong self-gravity, it will perturb their orbits, perhaps in a way that will contributes to the wave strength (e.g., \citealt{sh73}). Alternately, interactions between core eccentric orbits and other stars may ultimately destroy the coherence of the former, and destroy the wave. 

None the less, while they maintain their coherence, the core eccentric orbits provide a skeleton for realizing a stationary spiral structure like that of the Lin-Shu density wave theory (\citealt{li64}, \citealt{bt08}, Sec. 6.1), if not in the way originally envisioned. The waves of Fig. 6, existing mostly within the ILR, depend on eccentric orbits, and extend inward to a degree dependent on the most eccentric orbits. If individually observable, such orbits might not be readily recognized as belonging to the disc population at all. 

Finally, it is worth noting that if the most eccentric orbits were not be offset in azimuth by the disturbance that produced them, then they might form an inner bar that continues seamlessly into a spiral farther out. Both would have the same pattern speed.

\subsection{Spirals from other resonances}

It is also of interest to look at ensembles of orbits of varying eccentricity and sequentially offset, but originating from other resonances besides the ILR. Fig. 7 provides two examples. The top panel shows a case with orbits originating in the OLR Lindblad Zone. These are all very eccentric orbits, with $e$ varying between $0.8$ and $0.99$, and successively offset by a small angle of $\pi/128$. Given the small eccentricity range these orbits originate in a small patch just outside the OLR. Thus, the initial conditions are somewhat less specialized than the previous case, though fewer stars are included. In both cases the pattern speed was the same as in previous sections. 

\begin{figure}
\centerline{
\includegraphics[scale=0.5]{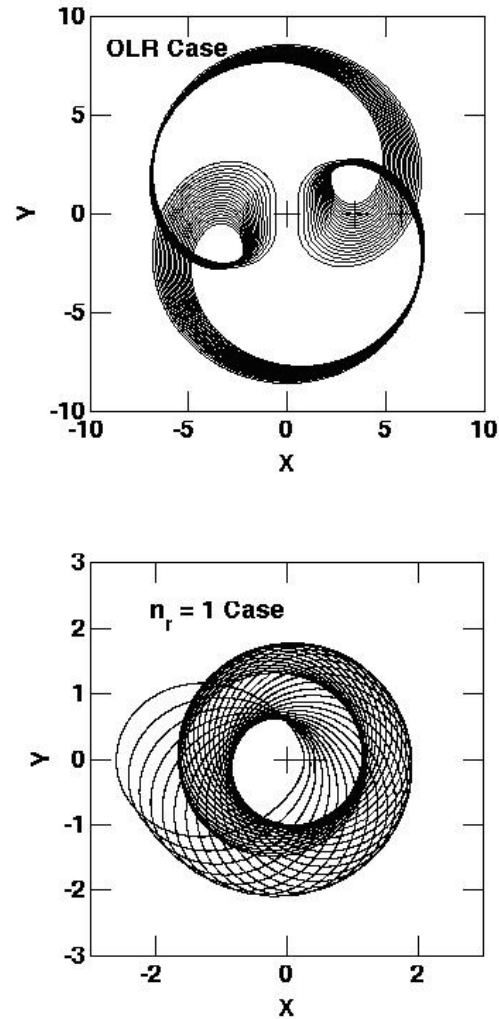}}
\caption{As in Fig. 6, spiral waves derived from sequentially offset resonant orbits, but here the orbits are derived from the OLR zone in the top panel, and from the inner $n_r = 1$ resonance in the bottom panel. In the top panel the eccentricities range from $0.8$ to $0.99$ in increments of $0.01$, and the azimuthal offset between successive orbits is $\pi/128$. In the right panel the eccentricities range from $0.3$ to $0.99$ in increments of $0.02$, and the azimuthal offset between successive orbits is $\pi/16$. Crosses mark the centre, corotation radius and OLR in the top panel, and the centre and $n_r = 1$ resonant radius on the bottom panel.}
\end{figure}

This orbit ensemble produces a prominent spiral, but unlike those of Fig. 6 the spiral does not simply fade as winds it outward. Instead, after passing through a pinch point its width steadily expands. Perhaps in some galaxies, without the compression of gas clouds, and the production of new stars, this outer expanded part is not visible. On the other hand, spirals that bend back toward the centre are visible in a number of galaxies, e.g., the Hubble Atlas galaxies NGC 718 and NGC 3504. Most examples are found in outer discs, as expected for OLRs. Most are also found in barred galaxies, perhaps excited by the bar. 

The outer parts of the orbits in this case look nearly circular. As noted previously, the inner parts of that curve might oppose an ILR derived bar. Alternately, they might generate their own bar, possibly in cooperation with orbits excited from an $n_r = 1$ resonance, like those shown in the bottom panel of Fig. 7 (but without the sequential offsets). If so, then again a bar-spiral combination, with a common pattern speed, would be a natural outcome. 

Other realizations with OLR resonant orbits, e.g., with larger azimuthal offsets between successive orbits, do not show such an extended spiral. Rather, they show compact regions around the pinch points, which increasingly trail the x-axis with greater azimuthal offsets. These might provide an explanation for the arcs or ansae \citep{bu13} seen outside and somewhat offset from the bars in some galaxies, e.g., NGC 5383. Outer rings could also be generated in this way if OLR orbits of modest eccentricity were excited over a range of azimuths. NGC 2217 is a possible example. The distinction here from classical resonant rings is minor, only that their orbits might be more eccentric than expected in the epicyclic theory. If the bar self-gravity grows these orbits may also ultimately become invariant manifold orbits \citep{ro12}.

The bottom panel of Fig. 7 shows an ensemble of orbits excited from the $n_r = 1$ resonance with $e$ values ranging from $0.3$ to $0.99$, with each separated by an $e$ increment of $0.02$. The azimuthal offset between each is $\pi/16$. Each of these orbits is an ellipse-like oval, but offset from the disc centre. In Fig. 7 they overlap to produce a very interesting wave set. An inner spiral winds outward counterclockwise until it pinches to a cusp. Over much of its length this inner spiral also forms the inner edge of an outer spiral that winds clockwise. The latter originates at a cusp and expands as it winds outward. 

This is again a very specialized initial condition. However, this form too is observed. A prototypical example is NGC 4622 discovered by \citep{by89} and modelled in some detail by Howard and Byrd (e.g., \citealt{by93}). They propose, on the basis of numerical models, that this unusual pattern is the result of a retrograde encounter with a companion galaxy, and that the inner arm at least could be long-lived. Such retrograde encounters can excite odd modes, especially the $n_r = 1$ mode, in discs. If these waves are indeed recently produced by an interaction, then the many dust lanes and feathers visible in the Hubble Heritage image of this object could be the result of gas clouds trying to pursue eccentric orbits cutting through the waves. 

The bottom panel of Figure 7 was a serendipitous result. It was quite surprising to find that such a pattern could be `modeled' so simply. (Note, that in more recent work, Byrd et al. discover a third, innermost set of waves, propose the existence of two corotation radii, and thus suggest it is an even more complex system than previously thought \citep{by08}.) 

Like the ILR patterns, those in Fig. 7 are persistent in the pattern frame. The longevity would increase the probability of them being discovered on the sky.

To return to interaction generated waves, we have previously found (\citealt{st90}, also in the analytic models of \citealt{st12}) exotic caustic forms in the non-rotating frame. These, of course, are either transient or subject to windup like epicyclic spirals. Note, however, that cusp caustics are visible in Fig. 7, and even more complex ones can be produced from resonant orbits in the right circumstances (see Sec. 5.2 below). 

\section{Resonant structures in near-solid-body potentials}
\subsection{Very simple bars}

The results of the previous sections were confined to the case of a flat-rotation-curve, logarithmic potential. Consideration of another very different case is instructive. The solid-body case has $\delta = -1$, in equation \eqref{eq3}, but that potential has the disadvantage of having singular properties. Let us instead consider the case of $\delta = -0.75$, which yields similar behavior, but is more generic in not having perfectly solid-body rotation. Note that equation \eqref{eq9} for the flat-rotation-curve case was the result of a perfect integral. In the present case, the corresponding integral can be evaluated numerically. Specifically, the term $2\pi/(1-e^2)^{1/2}$ in equation \eqref{eq9} is replaced by the approximate factor,

\begin{equation}
I = 2\pi \left(1.0003 - 0.01110e + 0.01109e^2
+ 0.06594e^3 \right).
\label{eq14}
\end{equation}

\noindent Otherwise the analysis proceeds as above.

\begin{figure}
\centerline{
\includegraphics[scale=0.35]{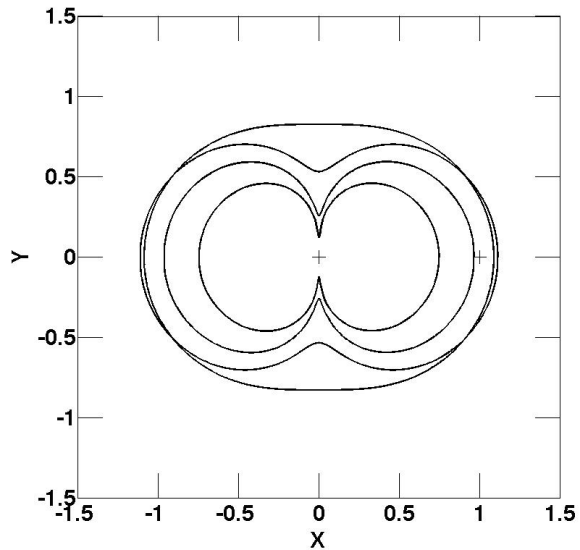}}
\caption{Several analytic, resonant eccentric orbits, as in Fig. 2, but here in the case of a near solid-body potential ($\delta = -0.75$). These are shown in the rotating pattern frame. The eccentricities, from the outer orbit inward, are $e = 0.5, 0.9, 0.99, 0.999$. }
\end{figure}

Fig. 8 shows a few ILR resonant orbits in this case, with eccentricities of $e = 0.5, 0.9, 0.99$ and $0.999$. At low eccentricity these orbits are in a lozenge-like shape. The two pinches are only significant at very high eccentricities. These latter cases are the equivalent of nearly radial orbits in this potential. Most of the orbits look rather `lozenge-like,' a perfect shape for a broad bar. Moreover, only for highly eccentric orbits does the outer radius change much. This is because it depends on a factor of $(1-e)^{-1/4}$, i.e., very weakly on $e$ unless $e$ is nearly equal to $1.0$. The exponent is even smaller for cases closer to the solid-body one, see equation \eqref{eq3}. Thus, in near-solid-body cases, the core structure of bars can be essentially a hollow lozenge. 

The lozenge shell appearance suggests a very simple analytic model in such cases. Self-gravity effects on the shell itself would be small, and the gravitational force on the interior and exterior could be computed in the thin shell limit, i.e., over a modest range of eccentricities. 

Beyond their simplicity another advantage of bars in these potentials is that a wide range of eccentricities can be excited in a small region. Alternately, over a wide range of eccentricity the orbits are nearly the same. This another example of why nature finds it easy to excite bars in rising rotation curve regions.

\subsection{Resonant rings and broad spirals}

As in Section 4.1, we can look at an ensemble of orbits of different eccentricity, with each successive orbit rotated by a small amount in azimuth. Fig. 9 shows 2 examples; the top panel includes orbits with $e$ ranging from $0.5$ to $0.99$. Two additional orbits were added with $e = 0.995$ and $0.999$. The azimuthal offset is $\pi/48$. The bottom panel is exactly the same except all eccentricities greater than $0.9$ were omitted. 

\begin{figure}
\centerline{
\includegraphics[scale=0.5]{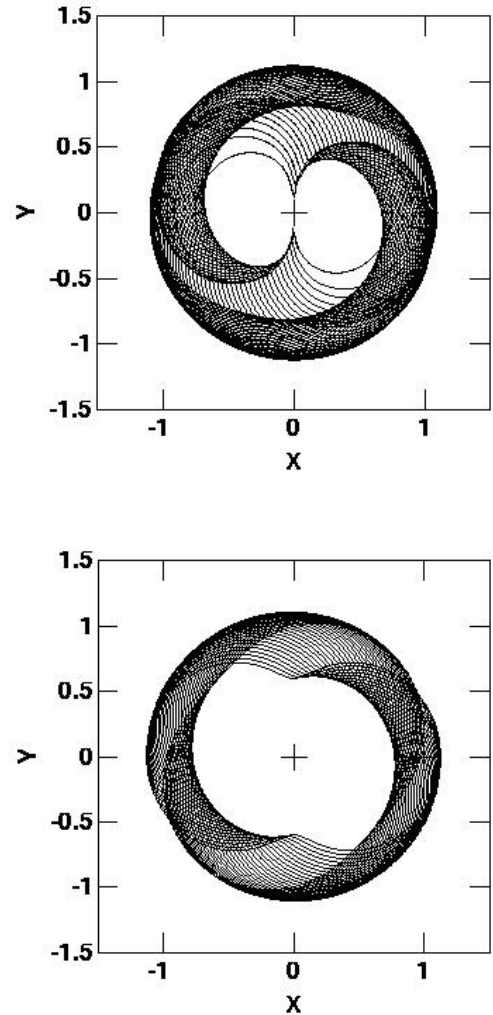}}
\caption{Spiral and ring examples in the near solid-body potential ($\delta = -0.75$). The azimuthal offset between successive orbits is $\pi/48$ in both panels. In the top panel the eccentricities range from $0.5$ to $0.99$ in increments of $0.01$, and in the bottom panel the eccentricities range from $0.5$ to $0.85$ also in increments of $0.01$. Crosses mark the centre, and the ILR in both panels. }
\end{figure}

The top panel shows a broad spiral, that pinches off as it is traced outward, like its FRC counterpart. However, it joins its opposite half before pinching off entirely. The spirals are also not as confused by numerous crossing orbit segments as in the FRC case. The overall impression is of spirals merging into a ring.

The comparison to the bottom panel shows that the inner spirals consist of very eccentric orbits, and so, probably a small fraction of the stars unless the initial disturbance produced mostly very eccentric orbits. Not surprisingly, the ring consists of less eccentric orbits. Swallowtail caustics of alternating orientation can be discerned within the ring in the bottom panel. The top panel shows that the inner spirals are extensions of two of these swallowtails. There are many orbit crossings within both spirals and ring. Gas clouds would collide and be shocked within these structures, perhaps inducing star formation. 

As previously, once formed, these are persistent structures, subject to disturbances. Analogous to the discussion in the previous subsection, the orbits making up the ring derive from a small radial range, but here from a region of azimuth varying somewhat with radius. 

Figs. 8 and 9 represent a small foray into the study of resonant orbits in nearly solid-body potentials, but the results are very promising. 

\section{Broad disturbances across multiple Lindblad Zones}
\subsection{Wind-up spirals, and proto-bulges}

In the preceding sections we considered what types of disc structures might be created by exciting the resonant eccentric orbits in individual Lindblad Zones. In this section we consider how broader disturbances, not tuned to the excitation pattern of any one zone, might trigger resonances in ranges of different Lindblad Zones.  Specifically, consider a disturbance that produces a constant azimuthal impulse $\Delta v_{\phi}$, for a significant radial distance along a line of constant azimuth, $\phi_d$. Also assume for simplicity that the radial impulse  $\Delta v_r$ is zero. This is like an impulsive tidal disturbance (see Sec. 2.2). 

Then, using equation (B1) from Paper 1, we can derive the following expression for the ratio of the specific angular momentum of a star before and after the disturbance,

\begin{equation}
\frac{h'}{h}  = 1+ \frac{\Delta v_{\phi}}{v_{\phi}} 
= \left( \frac{p}{r} \right)^{2(1-\delta)}.
\label{eq16}
\end{equation}

\noindent If $\Delta v_{\phi}, v_{\phi}$ are both constant (FRC, $\delta = 0$ case), then $p \propto r$ for excitations at different radii. In this case, the p-ellipse equation \eqref{eq3} gives,

\begin{equation}
\left( \frac{p}{r} \right)
 = \left[ 1+ e \cos \left( m(e) \phi_d \right) \right]^{1/2+\delta}.
\label{eq17}
\end{equation}

\noindent Since $p/r$ and $\phi_d$ are constant in the FRC case, then the eccentricity, $e$, and the value of $m(e)$ of the disturbed orbits are also constant with radius. Thus, over the relevant radial range, this constant azimuthal velocity impulse excites a line of p-ellipse resonances with the same value of $e$. At each of these radii we can view the p-ellipse as representing a pattern whose frequency equals the p-ellipse precession frequency, and the orbit is a symmetric ellipse in the frame of that local pattern. However, since $m(e)$ is constant, the local pattern speeds vary with radius as, $\Omega_p \propto \Omega \propto 1/r$. Thus, the global pattern, which is initially one of similar, concentric ellipses, will wind-up like a classical spiral density wave. 

Indeed, these orbits soon appear as a spiral. If the disturbance is substantial so that the eccentricity $e$ is large, then the stars will spend most of their time at the maximum radius. According to equation C1 of Paper 1, traversing a given angle at maximum radius will take a time of about $[(1+e)/(1-e)]^{1+2\delta}$ times what it would take near minimum radius. The spiral would consist of stars essentially `pausing' near maximum radius. These stars will initially be located along the radial line at azimuth $\phi_d$, but at each successive approach to their maximum radius, they will have sheared into a more tightly wound spiral form. Stars with the largest initial radii do not quite reach their maximum radii in this time interval.

\begin{figure}
\centerline{
\includegraphics[scale=0.6]{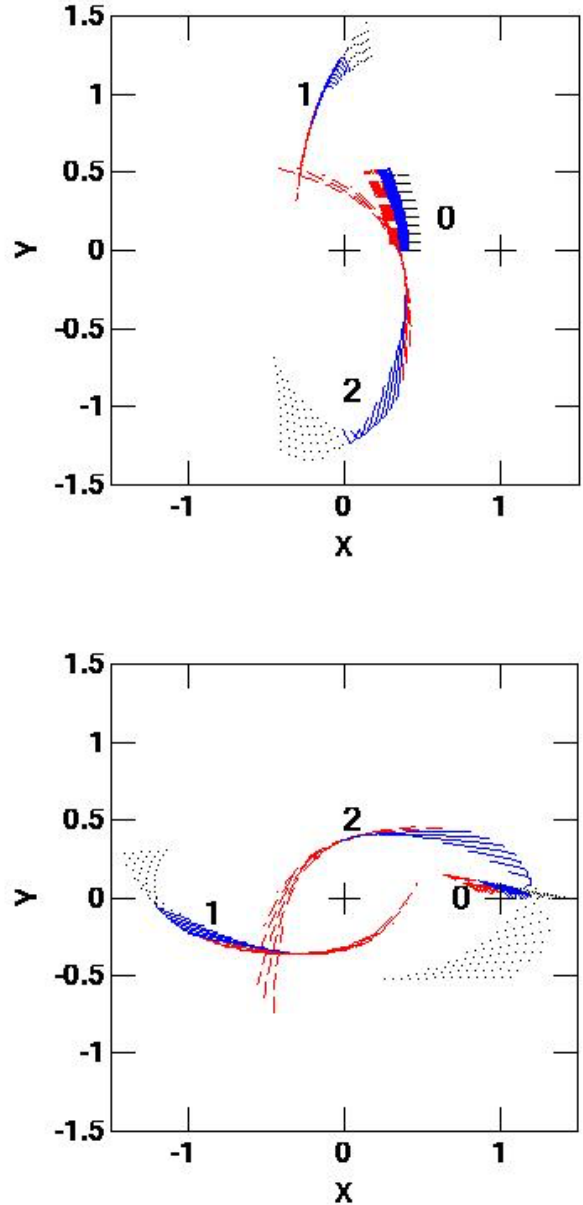}}
\caption{The top panel shows segments of 24 stellar orbital trajectories launched onto $e = 0.8$ p-ellipse orbits from their minimum radius by a positive azimuthal velocity impulse. The orbits with the lowest initial radii are shown as red, dashed curves, the outermost orbits as black dotted curves, and the middle orbits as solid blue curves. The three sets of trajectory segments correspond to time intervals: shortly after their impulsive launch (0),  around the approach to maximum radius by the middle (blue solid curve) stars (1), and around the time of the next approach to maximum radius by these stars (2). Crosses mark the centre of the disc and the radius of the classical ILR with the pattern speed used in previous sections. The bottom panel is the same except each trajectory is launched from its maximum radius following a negative azimuthal velocity impulse. The orbit segment in both panels are shown in the same rotating frame as previous figures, e.g., Figs. 2, 4-6. See text for further details.}
\end{figure}

This is illustrated in the top panel of  Fig. 10, which shows segments of a set of 24 orbits that all begin on the x-axis with a positive azimuthal velocity impulse that launches them outward onto p-ellipse orbits with eccentricity $e = 0.8$. The segments marked with a zero show the orbital progress for a short time after launch, and it is evident that the stars initially travel in parallel. The segments marked `1' show an interval around the time when the middle trajectories reach maximum radius. The stars that started at the largest radii (dotted black curves) have not yet reached their maximum radii. The stars that do reach maximum (solid blue, and dashed red curves) cluster on a curve spiral segment that has already turned significantly away from the initial radial locus. The time interval marked `2' occurs around the next maximum for the middle stars, as can be seen by the trajectory reversals in the figure. Again the dotted curves are either just or not quite reaching maximum. Most of the curves have left their maximum and are proceeding to the next. It is clear that the spiral segment around the maximum radii is now more nearly azimuthal than radial. That is, it has already wound up substantially, while the individual orbits are getting increasingly farther out of phase. 

The view in Fig. 10 is in the frame rotating at frequency which matches the precession of stars in the centre of the radial range shown. In the present case there is nothing special about this frame, since eccentric orbits are assumed to be excited over a significant radial range. With the assumed perturbation there is no unique resonant radius. 

Eventually, shear and windup will diminish the waves into invisibility. Gas clouds may participate in the first cycle. Recall, however, that a large fraction of the disc stars and clouds are on circular or epicyclic orbits, so clouds launched on very eccentric orbits will experience many collisions and be circularized and/or destroyed. 

The bottom panel of Fig. 10 is similar, except in the smaller area, we have more of an ocular wave (\citealt{el91}, \citealt{el00}), especially when we consider the symmetric wave originating on the opposite side of the disc and not shown in the figure. The combination of radial and azimuthal streaming motions evident in the figure clearly illustrate the complex motions within such a wave. 

Finally, if the disturbance is external, it is likely that there is also a perturbation perpendicular to the original disc plane. Thus, these waves may evolve in a thickened disc. Moreover, the stars on these eccentric orbits are likely to experience enhanced rates of scattering by the gas clouds on more circular orbits. This is essentially a scenario for the secular production of a bulge, whose formation timescale is the vertical scattering timescale of the eccentric stars. This scenario seems especially relevant for stars that experience a negative azimuthal impulse and fall inward. The wound up or phase mixed waves in a thickened disc effectively constitute a proto-bulge. (The positively perturbed stars may remain part of a thick disc.) This `scattered ocular' scenario for secular bulge formation differs from the usual buckling bar scenario (see Sec. V.d of \citealt{se14} and references therein), if only because there is no true bar involved.

\subsection{Persistent waves in transition regions of the rotation curve}

The last case we will consider is like the previous, only without the assumption that the rotation curve is flat throughout. Rather, we assume that it is flat at large radii, but that it rises in the inner regions, with decreasing slope as radius increases until it joins the flat segment. The transition region is potentially very interesting. An easy way to investigate this region is via an illustrative construction using a simple approximate rotation curve consisting of three power-law segments. The outermost segment is flat ($\delta = 0$). Going inward, the next segment is slightly rising, take $\delta = -0.2$ as a specific example. The final, innermost segment is somewhat steeper, e.g., $\delta = -0.4$.

Specifically, the following form will be chosen for the example rotation curve,

\begin{equation}
\delta = \left\{ \begin{array} 
{r@{\quad for \quad}c}
0.0\ (m_0 = \sqrt{2}) & r > 0.8\\
-0.2\ (m_0 = \sqrt{2.4}) & 0.7 < r \le 0.8\\
-0.4\ (m_0 = \sqrt{2.8}) & r \le 0.7.
\end{array} \right.
\label{eq18}
\end{equation}

\noindent Next, we set $\Omega = 1.0$ at $r =1.0$, and assume a fixed pattern speed of $\Omega_p = 0.2929$, so the classical ILR is again at $r =1.0$, as in the earlier examples. As in the previous subsection we assume a disturbance that imparts a constant azimuthal velocity impulse in a region along a line of fixed azimuth. In the present case we will assume that this excites p-ellipse orbits of eccentricity $e = 0.9$. In the FRC, ILR zone, with the given pattern speed, this orbit will be excited at a radius of about $r = 0.86$. This is the largest of the three resonant orbits shown in Fig. 11. 

\begin{figure}
\centerline{
\includegraphics[scale=0.33]{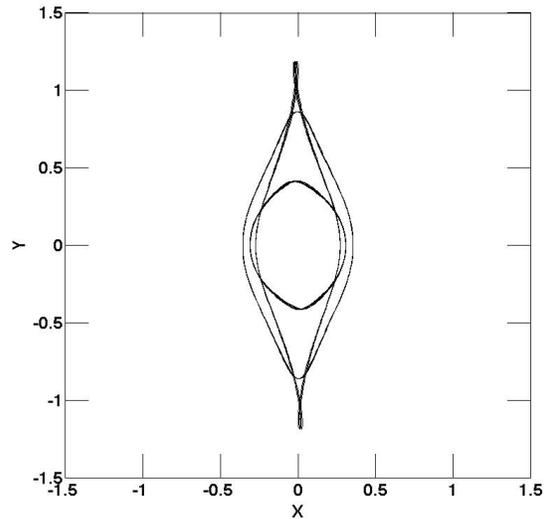}}
\caption{Three analytic, eccentric orbits derived from distinct eccentric ILRs in a disc with a radially varying rotation curve. The orbits are shown in a common rotating frame. See text for details.}
\end{figure}

In the middle segment of the rotation curve, the circular frequency increases as radius decreases but not as rapidly as in the FRC part. However, the ratio $m_o$ is larger in this segment than in the FRC, by nearly 10\%. These two factors offset each other in equations \eqref{eq1} or \eqref{eq2}. These trends continue into the innermost segment of the rotation curve. Thus, there are additional Lindblad Zones at the given pattern speed with resonant orbits at $r \simeq 0.75$ in the middle segment, and $r \simeq 0.46$ in the third segment, again with $e = 0.9$. Computing the value of the $p$ parameter (see equation \eqref{eq3}) required  numerical integrations of the differential equation for azimuth, like that in equation \eqref{eq15}. The results of these integrations, and fitting are,  

\begin{multline}
I = 2\pi\ exp \left(0.00636 - 0.252e + 1.895e^2 \right. \\
\left. - 3.535e^3 + 2.518e^4 \right), \ \ \delta = -0.2,
\label{eq19}
\end{multline}

\noindent and,

\begin{multline}
I = 2\pi\ exp \left(0.00103 - 0.0413e + 0.333e^2 \right. \\
\left. - 0.587e^3 + 0.429e^4 \right), \ \ \delta = -0.4.
\label{eq20}
\end{multline}

In fact, if the azimuthal impulse was of constant magnitude at all the relevant radii, equations \eqref{eq16} and \eqref{eq17} tell us that the eccentricities should increase somewhat with decreasing radii in the potentials of the innermost segments. This effect was neglected in producing Fig. 11, so the innermost orbits should actually be somewhat narrower. In any case, these orbits, and other near resonant orbits nearby would seem to provide very good seeds for growing a robust bar. The relative alignment of the orbits should be persistent in the rotating frame, subject to disturbances. 

The broader implications of this example are addressed in part by Fig. 12, which is a radius-frequency plot like Fig. 1, but changed to fit the rotation curve of equation \eqref{eq18}. The plot illustrates how, for the ILR zone, the steady increase of $\Omega$ with decreasing $r$ is offset by the abrupt reductions of $m_o$ (or $m_1$) in the segmented rotation curve. The plus signs show the $e = 0.9$ resonances of the different segments. It is also clear that the multiple excitation result is robust in the sense that the segment boundaries could be changed by quite a bit, and their $\delta$ values varied somewhat, without destroying the effect. Similarly, different velocity perturbation magnitudes exciting different radii in the Lindblad Zones could also be accomodated. 

The segmented rotation curve is an aid to separating the competing effects of a changing rotation curve on the circular and precession frequencies, but these effects would be the same in a smoothing changing, but convex, rotation curve region. There are limits to the effect. Firstly, if the region of change is small, then the number of stars excited onto resonant orbits may be too small to aid bar formation. If the change region is too broad and smooth, then the circular velocity increase with decreasing radius will dominate the precession effects, raising the Lindblad Zone above the pattern frequency for all but the most eccentric orbits. In this case, there would again be too few resonant orbits to have much effect. However, the example above, suggests there is a reasonable range of rotation curve convexity where the competing forces keep the moderate eccentricity part of the Lindblad Zone near the pattern frequency, and possibly oscillating above and below it, like the example. Thus, this appears to be a very viable means of triggering bar formation. 

In a way this phenomenon is not fundamentally different from classic models, where a change in the rotation curve in the inner regions allows the appearance of an 'inner' ILR in the epicyclic approximation. The primary difference is that the many eccentric orbits shown to form Lindblad Zones in the p-ellipse approximation, allow for a large variety of inner (eccentric) ILRs. 

\begin{figure}
\centerline{
\includegraphics[scale=0.35]{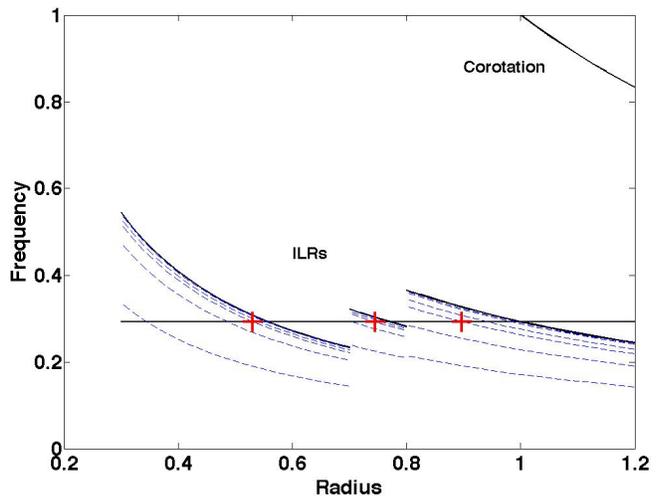}}
\caption{The structure of the Inner Lindblad Resonance Zones for the piecewise potential of equation \eqref{eq18}, along with the the pattern speed used to produce the orbits of equation \eqref{eq18}. The three red crosses mark the initial, excitation radii of the those orbits. Part of the corotation frequency curve is shown; compare to Fig. 1. See text for details.}
\end{figure}

\section{Summary and Outlook}

Previously (Papers 1 and 2) I have shown that p-ellipses approximate orbits very well in potentials like those in galaxy discs. Power-law potentials are described by the parameter $\delta$, orbit size by the parameter $p$, and the radial range by the parameter $e$ of equation \eqref{eq3} above. In Paper 2 it was shown that most residual deviations in orbit shape from the simple p-ellipse approximation can be fit with an additional harmonic term, and also by including the e-dependence of the precession rate, given by the parameter $m(e)$. This paper has focused on further ramifications of the $m(e)$ dependence, especially for our understanding of steady waves in galaxy discs. The fundamental point is that an eccentric orbit has a different value of precession than a nearly circular orbit, so it must have a compensating circular frequency (of its parent or guiding centre orbit) to be in resonance with a given (epicyclic) pattern frequency. This, in turn, means that the resonant radii are different for each value of the eccentricity, and each Lindblad resonance radius is actually a semi-infinite Lindblad Zone. 

The dependence of $m$ on $e$ is modest for most potentials of interest (e.g., equation \eqref{eq10}), and thus, a relatively minor correction to orbit approximations. The existence and properties of Lindblad Zones have more global ramifications.  Much of the analytic theory of bars and disc waves derives from the interaction of waves made of modestly perturbed epicyclic orbits with classical Lindblad resonances. Later these waves may evolve to nonlinear forms generally accessible to study only via numerical simulations. However, p-ellipse approximations, and the $m(e)$ relation, provide tools for partially filling the gap between near circular approximations and nonlinear numerical results. The existence of Lindblad Zones helps us understand direct excitation of resonant eccentric orbits by a strong perturbation.

These resonant, eccentric orbits could provide a backbone for a variety of waves. This backbone is a semi-material structure, which maintains its form in the pattern frame, while moving through the disc like a wave. Bars and spirals formed in this way could develop together in many cases. Several examples were considered in the previous sections, including the following.

1) In Sec. 3.3 the possibility of initiating bar formation by exciting resonant orbits in a sector of the ILR zone with a common pattern speed, but having a range of eccentricities was discussed. This is an extension of the same idea in epicyclic theory, and suffers the same problem - the resonant orbits would originate in a small region of the disc, so they would be a small population. Another difficulty for the eccentric orbits is that the impulsive disturbances assumed to excite them would have to vary strongly over the Lindblad Zone in order to excite multiple, proximate resonances of different eccentricity. This would require a moderate mass perturber located nearby. In essence, this process requires fine-tuning, and may not be likely.

2) In Sec. 4.1 a similar process was considered, with excitation along a curved locus in the ILR zone rather than a radial segment. This can generate some attractive spirals, that may be persistent, in the sense of not experiencing the usual wind-up. These waves also revealed interesting intersections with the orbits that produce them. However, fine-tuning of the initial conditions is also required in this case. 

3) In Sec. 4.2 it was shown that spirals can be produced in a similar way, at larger radii, in the OLR and $n_r = 1$ resonant zones. Again, fine-tuned initial conditions are required to produce the longest and tightest waves, but again, once produced these waves are persistent. In the OLR case, the excitation of a small range of eccentricities, in a small part of the Lindblad Zone could produce the arcs or ansae seen in many barred galaxies. The bar would be a natural excitation source, so this process seems more generic and probable than the previous cases. Moreover, rare galaxies can be produced by fine-tuned processes. Fig. 7 demonstrated that the $n_r = 1$ excitations are capable of generating an unusual wave set like that observed in NGC 4622. Although the initial conditions may be fine-tuned, the pattern persistence would make the discovery of such objects more likely. 

4) With no precession, it is relatively easy to generate persistent patterns in solid-body potentials. In Sec. 5, the question of to what degree near-solid-body potentials share this characteristic was considered. In Sec. 5.1 it was found that orbits excited in the ILR zone of a $\delta = -0.75$ potential form a lozenge-like shell, that would appear to provide a very good seed for bar development. In Sec. 5.2 it was shown that excitation on a curved locus can produce spirals like previous cases, but a more likely outcome is the generation of broad rings containing dense caustic regions.

5) In Sec. 6.1 we considered the well-known case of a constant velocity disturbance acting over a range of radii, and generating a spiral wave that winds up rapidly. A more novel aspect of this example is the fact that these waves consist of a dense orbit pile-up region at the maximum radius of quite eccentric orbits. It was noted that the disturbance that produced these eccentric orbits could have perturbed them out of a thin disc into a thick disc or proto-bulge. The turbulence generated by collisions between eccentric and more circular gas cloud orbits might feed the central nucleus, and facilitate central black hole growth. If so, this would provide a natural connection between secular bulge and central black hole growth due to the excitation of population of resonant eccentric orbits. 

6) In Sec 6.2 we investigated perhaps the most interesting case of all, a constant velocity impulse across a region where the rotation curve is changing from a rising form to a flat profile. In such a region the competing effects of changing circular and precession frequencies, together with the eccentricity dependence of the latter, can make the (ILR) Lindblad Zone oscillate through the pattern frequency at different radii. This can excite co-rotating, bar-like orbits over a significant radial range, thus, generating a larger population of such orbits than the previous cases. Given that this only requires a moderately convex portion of the rotation curve, and relatively probable initial conditions (e.g., from a large-scale tidal disturbance), this case seems generic. Depending on the excitation locus, persistent spirals could also be generated, as in previous cases. 

All of these kinematic waves develop and propagate in a purely symmetric gravitational potential. Many of them originate in a small fraction of the disc, so they are few and may be relatively fragile. However, self-gravitational effects, and especially the entrainment of nearby orbits, might build them into stronger waves. The relation to very nonlinear gravitational processes, such as swing amplification, is not clear. 

Though the less regular orbits might be more easily scattered out of the wave, and gas clouds destroyed by feedback effects, some fraction of the backbone orbits may retain their coherence to generate additional cycles of wave activity. This conjectural process is different than that recently proposed by \citep{se14} for the production of recurrent wave patterns, but could well be compatible with it.

There are many more cases not explored in the previous sections. The relationship between transient and resonant waves excited by external disturbances certainly deserves further study. It should also be possible to extend the models above to weakly self-gravitating cases that might come closer to connecting with simulation results. This will be pursued in a future work. Conversely, many numerical simulations to test the role of eccentric orbits in disc waves suggest themselves. Overall, the `missing link' of eccentric orbits and their Lindblad Zones, which can be studied with p-ellipse approximations, may provide insight into many questions in galaxy disc dynamics.

\section*{Acknowledgments}

I am grateful for a number of helpful comments and corrections from the referee, G. G. Byrd, and for inputs from Beverly Smith, S. Ram Valluri, and Bruce Elmegreen. I acknowledge use of NASA's Astrophysics Data System, and the NASA Extragalactic Data System.

\bibliographystyle{mn2e}

\bsp
\label{lastpage}
\end{document}